\documentclass[twocolumn]{article}

\usepackage[margin=0.7in]{geometry}
\usepackage[utf8]{inputenc} 
\usepackage[T1]{fontenc}    
\usepackage{lmodern}
\usepackage{hyperref}       
\usepackage{url}            
\usepackage{booktabs}       
\usepackage{amsfonts}       
\usepackage{nicefrac}       
\usepackage{microtype}      
\usepackage{graphicx}
\usepackage{subcaption}
\usepackage{color}
\usepackage{amsmath}
\usepackage{todonotes}
\usepackage{dsfont}
\usepackage{multirow}

\begin{document}


\title{Collaborative Filtering with Recurrent Neural Networks}
%
%
%
%
%

\author{
%
%
    \textbf{Robin Devooght}\\
    IRIDIA\\
    Universit\'{e} Libre de Bruxelles\\
    1050 Brussels, Belgium\\
    \texttt{robin.devooght@ulb.ac.be}
    \and
    \textbf{Hugues Bersini}\\
    IRIDIA\\
    Universit\'{e} Libre de Bruxelles\\
    1050 Brussels, Belgium\\
    \texttt{bersini@ulb.ac.be}
}

\date{}

\maketitle
\section*{Abstract}
We show that collaborative filtering can be viewed as a sequence prediction problem, and that given this interpretation, recurrent neural networks offer very competitive approach.
In particular we study how the long short-term memory (LSTM) can be applied to collaborative filtering, and how it compares to standard nearest neighbors and matrix factorization methods on movie recommendation.
We show that the LSTM is competitive in all aspects, and largely outperforms other methods in terms of item coverage and short term predictions.

\paragraph{Keywords.} Collaborative filtering, recommendation systems, recurrent neural network, LSTM, deep learning.

\section{Introduction}

Collaborative filtering is the problem of recommending items to users based on past interactions between users and items.
The intuition is that if two users have had similar interactions in the past, they should look to each other for recommendations.
The nearest neighbors approaches (KNN) are based on that idea, and offer multiple mechanisms to identify similar users and pool recommendations.
Next to KNN, the other main approach for collaborative filtering is matrix factorization (MF), which frames recommendation as a dimensionality reduction problem.
MF attempts to represents users and items as points in a feature space according to an optimization criterion based on the interactions between users and items (i.e. the users should be close to the items that they liked).
Although useful, those methods are far from perfect and improvements are coming more and more slowly. 
Moreover, they are unadapted to capture the temporal aspects of recommendations, such as evolving taste or context-dependent interests.

We explore a new approach to collaborative filtering, based on recurrent neural networks.
Modern recurrent neural networks such as the LSTM are powerful tools for sequence prediction problems and are well-suited to capture the evolution of users taste.
In order to apply them to recommendations, we need to reframe collaborative filtering as a sequence prediction problem.

In the following sections, we will describe collaborative filtering as a sequence prediction problem, then show how the LSTM can be trained on such a problem and compare it to traditional collaborative filtering approaches.
We will then explore some of the many design choices of the LSTM.

Our code is available in \url{github.com/rdevooght/sequence-based-recommendations}.

\section{Collaborative filtering as a sequence prediction problem}
\label{sec:CF_as_SPP}

In a typical top-N recommendation problem, we consider a user $i$ at a time $t$, with $\mathcal{S}_{i}^{t^{-}}$ being the set of items that 
the user consumed before $t$ and $\mathcal{S}_{i}^{t^{+}}$ the set of items that he consumed after $t$. The goal of the recommendation system is to predict the items in $\mathcal{S}_{i}^{t^{+}}$ as a function of $\mathcal{S}_{i}^{t^{-}}$.
In this setting, that we call \emph{static}, the order in which items are consumed is irrelevant for the recommendation system.

Instead, we suggest that the recommendations should be based not only on the set of items that the user consumed, but on the order in which he consumed them.
The user is represented by its sequence of action: he consumed $x_1$, then $x_2$, then $x_3$, and the goal is to predict his next actions ($x_4$, $x_5$, etc.) based on the beginning of the sequence.

Framing collaborative as a sequence prediction problem naturally leads to a distinction between what we call short-term and long-term predictions.
A short-term prediction aims to predict which item will the user consume \emph{next} (i.e. right after the last one), while a long term prediction aims to predict which items will the user consume \emph{eventually}.
In the static setting, this distinction does not make sense because the order of items in $\mathcal{S}_{i}^{t^{+}}$ is ignored, and predictions in a static setting are equivalent to long term prediction.
In sequence prediction however, the models are usually trained to produce short-term prediction.

It is hard to argue for which type of prediction is more interesting in recommender systems, and ultimately, this choice lies with the user, but it is useful to remember that a sequence prediction approach is better suited for short-term predictions.
Moreover, we will show in Section \ref{sec:diversity} that training for short-prediction is likely to improve the diversity of recommendations.
In the following experiments, we will report performances on both short-term and long-term predictions.

The basic motivation supporting this work is that, while neglected so far, the information contained in the sequence of action could be of great importance for producing better recommendations.
For example, this sequence can reveal the evolution of a user's taste. It might help to identify which items became irrelevant with regards to the current user's interest, or which items make part of a vanishing interest.
It might also help to identify which items are more influential in changing users taste.

Methods based on sequence prediction should produce richer models. As a matter of fact, a static recommendation system models users and items as points in a feature space, frozen in time. It will recommend items that are close to the user in that specific poorly informative space. Given a sequence prediction approach, users are rather represented by trajectories in the feature space, entailing a more accurate prediction of the evolution of the users interest (because pursuing this same trajectory). 

\section{Sequence prediction favors diversity in recommendations}
\label{sec:diversity}

As said in the previous section, methods that ignore the sequence of events are trained to produce long-term predictions, while methods based on sequence prediction are usually trained to produce short-term predictions.
The typical metrics to evaluate long term predictions are the precision and recall.
If $\mathcal{P}_{i}$ are the predictions based on $\mathcal{S}_{i}^{t^{-}}$, then the precision is $|\mathcal{P}_{i} \cap \mathcal{S}_{i}^{t^{+}}| / |\mathcal{P}_{i}|$ and the recall is $|\mathcal{P}_{i} \cap \mathcal{S}_{i}^{t^{+}}| / |\mathcal{S}_{i}^{t^{+}}|$ (we usually compute the precision and recall ``at $k$'', where $k |\mathcal{P}_{i}|$).
In order to evaluate the quality of short-term prediction, we propose a simple metric that we will call the ``sequence prediction success at $k$'' ($\text{sps}@k$) : $|\mathcal{P}_{i} \cap \{x_{i}\}|$, where $x_i$ is the first item of the sequence $\mathcal{S}_{i}^{t^{+}}$.

An interesting side-effect of training recommendation systems to optimize short-term rather than long term predictions, is that it increases the diversity of the recommendations.
Although we do not have a formal proof, the reasoning is as follows: the correct short term predictions are a subset of the correct long term predictions; 
because of that, any given item will be a correct prediction for more users in terms of long term prediction than in terms of short term prediction; 
the result is that it takes fewer items to make correct long term predictions for a given percentage of users than it takes to make correct short term predictions.

\begin{figure*}
  \centering
  \begin{subfigure}{0.45\textwidth}
        \centering
        \includegraphics[scale=0.95]{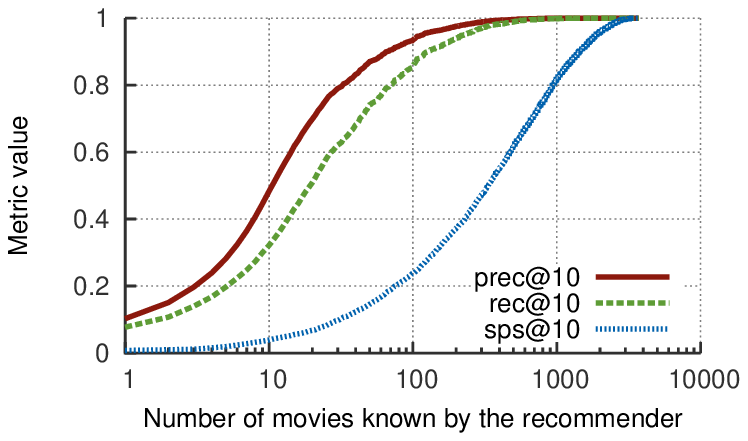}
        \caption{Movielens 1M}
    \end{subfigure}
    \begin{subfigure}{0.45\textwidth}
        \centering
        \includegraphics[scale=0.95]{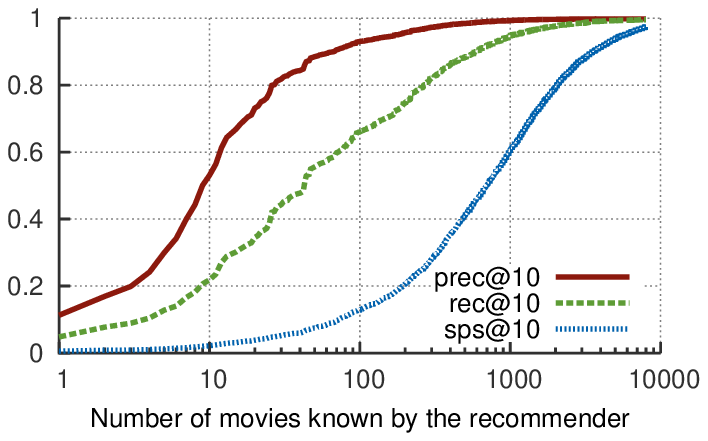}
        \caption{Netflix}
    \end{subfigure}
  \caption{Evolution of the $\text{rec}@10$, $\text{prec}@10$ and $\text{sps}@10$ for a recommendation that can only recommend the top $t$ most popular items.
  The maximum value of the $\text{rec}@10$ is actually much smaller than 1 but we normalized the recall here by dividing it by its maximum value to make the graph more readable.}
  \label{fig:recommendation_diversity}
\end{figure*}

We illustrate that with a simple experiment. Consider an oracle (i.e. a perfect recommendation system) that can only recommend items within the $t$ most popular ones; Figure \ref{fig:recommendation_diversity} shows how the $\text{rec}@10$, $\text{prec}@10$ and $\text{sps}@10$ increase as $t$ increases on two real datasets.
The $\text{rec}@10$ and $\text{prec}@10$ converge very fast, they reach 80\% of their maximum value with a small fraction of the items, and each new item brings only a marginal improvement. The $\text{sps}@10$ on the other hand has a much slower convergence and requires therefore a higher diversity of recommendation to reach 80\% of its maximum value; we therefore expect that optimizing a recommendation system for short term prediction will force it to produce more diverse recommendations.

\section{Methods comparison}
\label{sec:experiments}

Recurrent neural networks (RNNs) might be the most versatile methods of sequence prediction, and their recent success in speech recognition, translation and other domains \cite{graves2013speech, sutskever2014sequence} make them good candidates for the recommendation problem.
In this section we evaluate RNN on a problem of item recommendation for two movie recommendation datasets, and compare it to collaborative filtering methods (that makes no use of the sequence information).

\subsection{Datasets}
\label{sec:datasets}

Our major constraint in finding datasets is the presence of information on the order of events (often in the form of a timestamp).
This information is unfortunately missing in most collaborative filtering dataset, but is available in two well known datasets of movie recommendation:

\begin{itemize}
    \item Movielens 1M: a rather small version of the Movielens dataset, with 6040 users and 1,000,209 ratings over 3706 movies.
    This dataset has extra information about the users (age, sex, occupation) and about the movies (year, genre) that will be exploited in Section \ref{sec:rnn_parameters}.
    \item Netflix: A much larger dataset, with about 480k users and 100M ratings over 17770 movies.
\end{itemize}

Each set has been splitted into training, validation and test subsets.
These subsets where obtained by dividing the users into 3 groups: $N$ randomly chosen users and all their ratings to constitute the test set, $N$ others to constitute the validation set and all the remaining users for the training set.
We used $N = 500$ for Movielens 1M and $N = 1000$ for Netflix.

In those datasets, any item can appear only once in the rating history of any user. We therefore helped all the methods by forcing them to recommend items that the user had not yet seen.

Although the datasets have explicit feedback of users in the form of ratings, none of the methods presented here used the values of the rating to build their model or make predictions, they just used the fact that an item was rated or not by a user.

\subsection{Recurrent neural networks}

RNNs are commonly used for language modeling, where they are trained to learn sequences of words \cite{mikolov2010recurrent}.
We took a similar approach by considering each item as a word, the catalog of items as the full vocabulary, and the history of each user as a sample sequence.
The RNN runs through the sequence of items consumed by a user, item by item.
At each time step, the input is the one-hot encoding of the current item, and the output is a softmax layer with a neuron for each item in the catalog.
The $k$ items whose neurons are activated the most are used as the $k$ recommendations.

The state-of-the-art in recurrent neural networks is what are called ``gated'' RNNs, where the internal state of the RNN is controled by one or more small neural networks called gates.
The original gated RNN is the LSTM\cite{LSTM}, but it has spawned multiple variants \cite{GRU, greff2015lstm}.

We trained the RNN to minimize the categorical cross-entropy loss, with the only correct item being the next item in the sequence.
As said earlier, this approach trains the RNN to focus on short-term prediction so that we can unsurprisingly expect to have a high sps.

For the following experiments, we used a unidirectional single-layered LSTM, we tuned the number of hidden neurons and the learning mechanism over the validation set. 
The results shown here are obtained using Adagrad \cite{adagrad} with a learning rate of $0.1$, $20$ hidden neurons for the Movielens dataset and $100$ for the Netflix dataset.

\subsection{Competing methods}

We compare the RNN with two static methods of top-N recommendation: one based on nearest neighbors and one on matrix factorization.
We also compare the RNN with a simple Markov chain that can be seen as a baseline for the sequence prediction approach.

\subsubsection{Markov chain}

In this simple method, the users' behavior is modelled by a Markov chain whose states are the different items.
The transition probabilities between items are inferred from the transition frequencies observed in the training set.
At any time the state of a user corresponds to the last item that he consumed, and the recommendations for that user will be the $k$ items with the highest transition probabilities from that state.
In other words, if the last item consumed by a user is the item $j$, the $k$ first recommendations of the Markov model will be the $k$ items that followed most often the item $j$ in the sequences coming from other users.
This is equivalent to a bigram model in language modeling in which the words become the items.

\subsubsection{User-based nearest neighbors}

User-based nearest neighbors or user KNN is one of the oldest method of collaborative filtering, yet still a strong baseline for top-N recommendation.
A score $s_{ij}$ is computed between a user $i$ and an item $j$:
\begin{equation}
    s_{ij} = \sum_{u \in \mathcal{N}_k(i)} c_{iu} \mathds{1}(j \in \mathcal{S}_{u})
\end{equation}
Where $c_{iu}$ is the similarity between users $i$ and $u$, $\mathcal{N}_k(i)$ is the set $k$ users closest to $i$ according to the similarity measure $c$, and $\mathds{1}(j \in \mathcal{S}_u)$ is the indicator function that evaluates to 1 if item $j$ belongs to the sequence of items of user $u$, or else evaluates to 0.
We used the cosine similarity measure, which is usualy prefered for item recommendation:
\begin{equation}
    c_{iu} = |\mathcal{S}_i \cap \mathcal{S}_u| / \sqrt{|\mathcal{S}_i| |\mathcal{S}_u|}
\end{equation}
The size of the neighborhood ($k$) was optimized by means of a validation set.

\subsubsection{BPR-MF}

BPR-MF is a state-of-the-art matrix factorization method for top-N recommendation devised by \cite{BPR}.
It is based on the Bayesian personalized ranking: an objective function similar to the AUC (area under the ROC curve) that can be trained trough stochastic gradient descent.
We used the original implementation of BPR-MF, available in the MyMediaLite framework \cite{Gantner2011MyMediaLite}.
BPR-MF has many parameters; we selected the most important (number of features and number of iterations) on the basis of a validation set, and kept the default values of MyMediaLite for the others.

\subsection{Metrics}
\label{sec:metrics}

We compare the different methods through a range of metrics finely designed to capture various qualities of the recommendation systems.
\begin{itemize}
    \item \textbf{sps}. The Short-term Prediction Success captures the ability of the method to predict the next item. It is 1 if the next item is present in the recommendations, 0 else.
    \item \textbf{Recall}. The usual metrics for top-N recommendation captures the ability of the method to do long term predictions.
    \item \textbf{User coverage}. The fraction of users who received at least one correct recommendation. Average recall (and precision) hide the distribution of success among users. A high recall could still mean that many users do not receive any good recommendation. This metrics captures the generality of the method.
    \item \textbf{Item coverage}. The number of distinct items that were correctly recommended. It captures the capacity of the method to make diverse, successful, recommendations.
\end{itemize}
All those metrics are computed ``at 10'', i.e. in a setting where the recommendation systems produces ten recommendations for each user.

\subsection{Testing procedure}

For each user of the test set, the method can base its recommendations on the first half of the user's ratings and on the model previously built on the training set (or on the training set directly, in the case of the nearest neighbours methods).
Those recommendations are evaluated against the second half of the user's rating, using the metrics described in Section \ref{sec:metrics}.
The metrics are then averaged over all test users.

The only method that needs the ratings of a user during the construction of the model is BPR-MF.
For that reason, BPR-MF is trained on a larger training set than the other methods, that includes the first half of the ratings of each of the test users.
This is an unfair but unfortunately unavoidable advantage for the BPR-MF method.

Since BPR-MF and RNN produce stochastic models, table \ref{tab:results} gives the average and the standard deviation of the results over ten models.

\begin{table*}[t]
  \caption{Comparison of top-N recommendation methods on Movielens 1M and Netflix}
  \label{tab:results}
  \centering
  \begin{tabular}{clllll}
    \toprule
    & & \multicolumn{4}{c}{Metrics (@10)} \\
    \cmidrule{3-6}
    & Method & sps (\%) & Item coverage & User coverage (\%) & rec (\%) \\
    \midrule
    \multirow{4}{*}{\rotatebox[origin=c]{90}{Movielens}} & BPR-MF & $12.18 \pm 0.35$ & $388.4 \pm 7.91$ & $82.56 \pm 0.48$ & $5.64 \pm 0.07$   \\
    & User KNN & $14.40$ & 277 & $80.8$ & $6.31$   \\
    & MC & $29.20$ & 518 & $77.0$ & $4.90$   \\
    & RNN & $33.69 \pm 0.58$ & $649.22 \pm 7.88$ & $86.62 \pm 0.38$ & $7.63 \pm 0.1$ \\
    \midrule
    \multirow{4}{*}{\rotatebox[origin=c]{90}{Netflix}} & BPR-MF & $9.76 \pm 0.35$ & $589.22 \pm 7.13$ & $79.96 \pm 0.36$ & $6.92 \pm 0.09$ \\
    & User KNN & $13.04$ & $383$ & $80.84$ & $8.49$ \\
    & MC & $32.50$ & $594$ & $64.50$ & $3.17$ \\
    & RNN & $40.62 \pm 0.70$ & $769.56 \pm 8.91$ & $80.59 \pm 0.36$ & $6.10 \pm 0.10$ \\
    \bottomrule
  \end{tabular}
\end{table*}

\subsection{Analysis}

The results are shown in Table \ref{tab:results}.
The methods using the sequence information are impressively much better than the others in terms of sps.
It is worth underlying the quality of those results: given ten trials (i.e. ten recommendations), the RNN is able to predict the next movie seen by 33\% of the users of Movielens and 40\% of the users of Netflix, while methods not based on the sequence are below 15\%.
As predicted in Section \ref{sec:diversity}, methods with the best sps also have a larger item coverage, which is an important, but often overlooked aspect of recommendations.
In particular, the item coverage of the RNN is more than twice the one of the User KNN.

As a global observation, the RNN proves to be a very promising method for all considered metrics.
It dominates the other methods in every aspects on the Movielens dataset and is only beaten in terms of recall by BPR-MF and User KNN on Netflix.

\section{Variations of the recurrent neural networks}
\label{sec:rnn_parameters}

As we apply RNN to a new problem, it is worth exploring how to adapt it to the specificities of the new problem.
In this section we study technical details of the implementation such as the learning rate and the number of hidden neurons, then we present a modification of the loss function that leads to more diverse recommendations, and we study how the RNN can use extra information such as users' age or ratings value to built better models.

\subsection{Learning method}

Modern neural networks are rarely trained using the vanilla SGD approach: a set of mechanisms have been developped to speed up learning and schedule the decrease of the learning rate.
Two mechanisms have been found to be generally useful: 
\begin{itemize}
    \item Momentum: smooths the gradients variations over time in order to avoid ``zig-zags'' during learning\cite{momentum}.
    \item Adaptive learning: decrease the learning rate for frequently updated parameter (Adagrad\cite{adagrad}, rmsprop).
\end{itemize}
Some methods, such as Adadelta\cite{adadelta} and Adam\cite{adam} combine both approaches.

\begin{figure}
  \centering
    \includegraphics[scale=0.95]{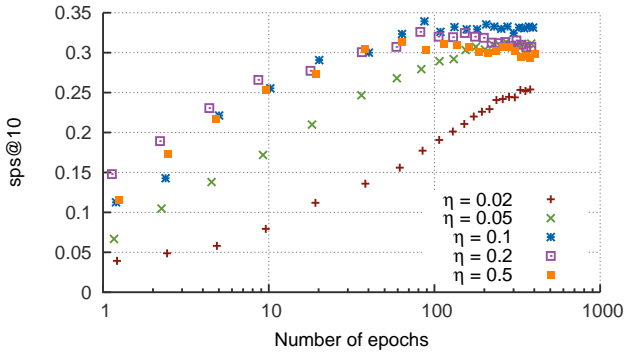}
  \caption{Influence of the learning rate parameter with the adagrad scheduler on Movielens. The sps is computed on the validation set during learning.}
  \label{fig:adagrad}
\end{figure}

We observed that in this context, adaptive learning seems more important than momentum, with the methods Adagrad, rmsprop and Adam working particularly well.
Adagrad is especially appealing because it requires to tune only one parameter: the initial learning rate, while rmsprop requires two, and Adam three.
The Adagrad updates are computed in the following way:
\begin{equation}
    \theta_{t+1} = \theta_{t} - \frac{\eta}{\sqrt{G_{t} + \epsilon}} \odot g_{t}
\end{equation}
Where $\theta$ is the vector of parameters, $\eta$ is the initial learning rate, $\epsilon$ is a small value to avoid division by zero, $g_{t}$ is the gradient of the objective function with regards to the parameters at time $t$, and $G_{t} = \sum_{i=0}^{t} g_{i}^{2}$ is the sum of previous square gradients.
All parameters start with the same learning rate, but during training the learning rate decreases faster for the parameters that are often associated with large gradients.
Those parameters will therefore be fixed early during training, making it easier to learn the other parameters, which have more subtle influences on the objective function.
Figure \ref{fig:adagrad} shows the influence of the learning rate with the Adagrad method.
Learning rates around $0.01$ gave the best result, and similar observations where made on the Netflix dataset.

\subsection{Influence of the cell size}

\begin{figure*}
  \centering
  \begin{subfigure}{0.45\textwidth}
        \centering
        \includegraphics[scale=0.95]{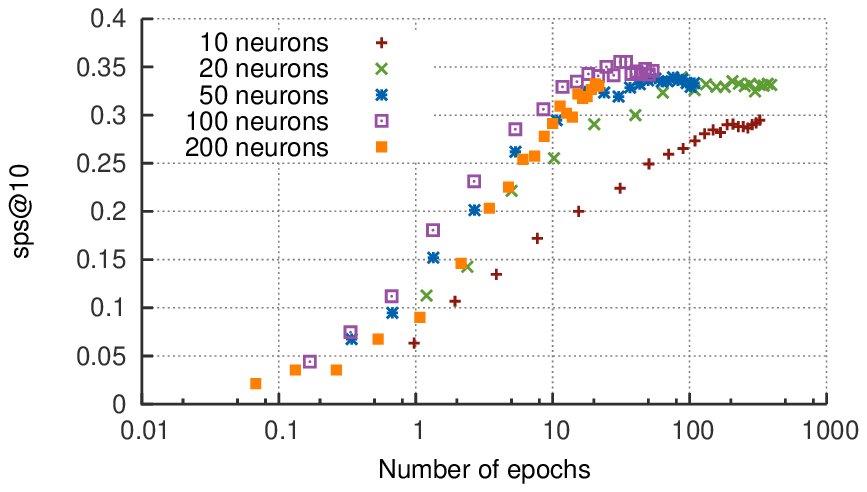}
    \end{subfigure}
    \begin{subfigure}{0.45\textwidth}
        \centering
        \includegraphics[scale=0.95]{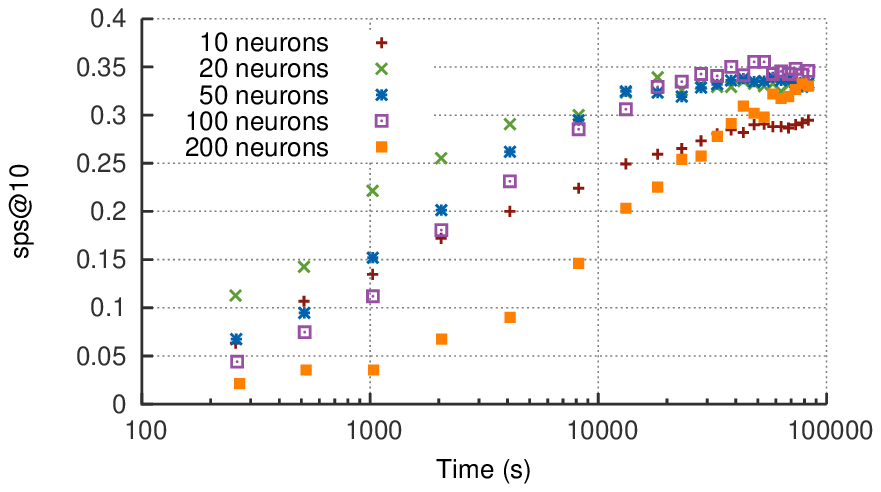}
    \end{subfigure}
  \caption{Influence of the number of neurons on Movielens. The sps is computed on the validation set during learning. The right figure shows the sps in function of the training time on one core of an Intel(R) Xeon(R) CPU E5-2620 v3 @ 2.40GHz.}
  \label{fig:neurons}
\end{figure*}

Another important parameter of the of the LSTM is the number of neurons in the cell. 
The LSTM will not be able too learn a rich enough model if the number of neurons is too small, but more neurons lead to a slower learning and it may increase the risk of over-fitting.
A good first choice for the number of neurons in the LSTM cell is the number of feature you would use to solve the same problem with matrix factorization.
Figure \ref{fig:neurons} shows the validation error during training for several cell size.
We observe that with 10 neurons the model is severely limited, and the highest sps seems to be reached with around 100 neurons.
However, we also observe a faster learning for the smaller cell of 20 neurons.

\subsection{Architecture}

Although we used a vanilla LSTM in all our experiments, many other RNN architectures could have been used.
In this section we very briefly explore some of the other possible types of RNNs:
\begin{itemize}
    \item Bidirectional LSTM: two LSTM are used in parallel, one reading the inputs in chronological order and the other in the reverse order. The output of both LSTM is fed into the last (softmax) output layer.
    \item 2-layered LSTM: RNN can be stacked, the output of one LSTM feeding the next one. We tried the simplest version, with two layers of LSTM, the first reading the initial input, and the second reading the output of the previous layer and producing the predictions.
    \item GRU: the gated recurrent unit is simpler than the LSTM, with fewer gates and fewer parameters to tune\cite{GRU}.
\end{itemize}

\begin{figure*}
  \centering
  \begin{subfigure}{0.45\textwidth}
        \centering
        \includegraphics[scale=0.95]{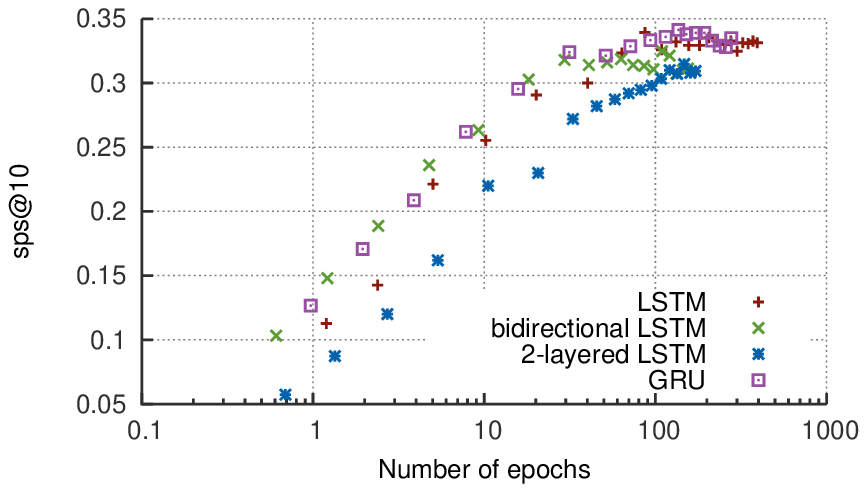}
    \end{subfigure}
    \begin{subfigure}{0.45\textwidth}
        \centering
        \includegraphics[scale=0.95]{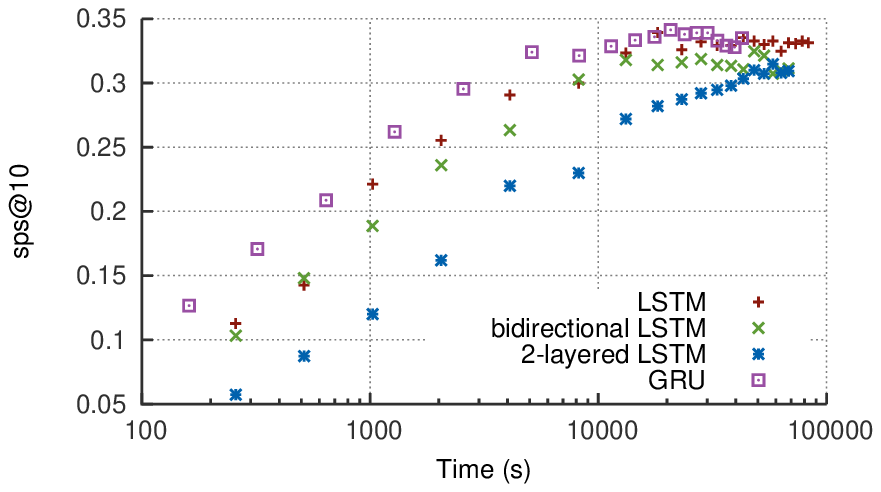}
    \end{subfigure}
  \caption{Influence of the architecture of the RNN on Movielens. Each type (LSTM, bidirectional LSTM, 2-layered LSTM and GRU) have a cell size of 20 neurons. The sps is computed on the validation set during learning. The right figure shows the sps in function of the training time on one core of an Intel(R) Xeon(R) CPU E5-2620 v3 @ 2.40GHz.}
  \label{fig:arch}
\end{figure*}

Figure \ref{fig:arch} compares the different architectures with the vanilla LSTM.
We report the evolution of the validation error during training, both we regards to the number of epochs and to the training time because the choice of architecture has a significant impact on the speed at which the network is able to process a sequence.
In particular, the GRU is faster than the LSTM-based RNNs, and the gain in speed does not seem to impact the quality of the prediction.
On the long run however, the difference between GRU and LSTM is negligible.
The bidirectional and the 2-layered LSTM slightly under-perform, but the overall effect of the choice of architecture appears to be small.

\subsection{Diversity bias}

As expressed earlier, diversity is a key challenge of recommendation systems.
The difficulty comes mainly from the fact that the distribution of popularity among items is usually very skewed, in the presence of very few popular items and much more rarer items.
Naturally, any model trained to optimize the chance of correct recommendations will learn to often propose the most popular items.
Unfortunately, most of the times, those popular items turn out to be trivial, useless recommendations.

We present here a small modification of the objective function of the RNN, aimed to increase the diversity of the recommendation while not loosing too much precision.
In Section \ref{sec:experiments} we used the categorical cross-entropy as the objective function of the RNN: $L = -\log(o_{\text{correct}})$, where $o_{\text{correct}}$ is the value of the output neuron corresponding to the correct item.
We test here a slightly different objective function that lowers the error associated with mispredicting the most popular items. The underlying rational is that insisting on the rarest items should counteract the bias toward popular items caused by the skewed distribution of popularity.

Our objective function is thus transformed as follows:

\begin{equation}
    L_{\delta} = -\log(o_{\text{correct}}) / e^{\delta p_{\text{correct}}}
\end{equation}

Where $\delta \in [0, \inf)$ is the diversity bias parameter and $p_{\text{correct}}$ is a measure of popularity associated with the correct item.
$p$ could be for example the number of views or the number of purchases of the item.
In our experiment, we constructed $p$ by dividing the items into ten bins of logarithmic size, the smaller bin containing the most popular items and the largest bin containing the least popular (in terms of number of ratings). Then we set $p = 1$ for all the items in the largest bin, $p = 2$ for the items in the second largest, etc.
When $\delta = 0$, the objective function is reduced to the categorical cross-entropy, and increasing $\delta$ increases the bias towards infrequent items.

\begin{figure}
  \centering
    \begin{subfigure}{0.45\textwidth}
        \centering
        \includegraphics[scale=0.85]{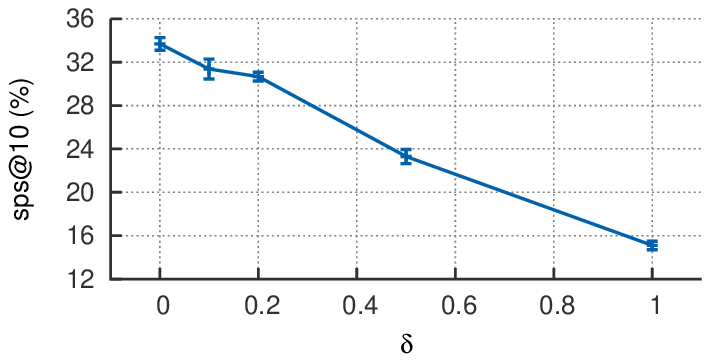}
    \end{subfigure}
    \begin{subfigure}{0.45\textwidth}
        \centering
        \includegraphics[scale=0.85]{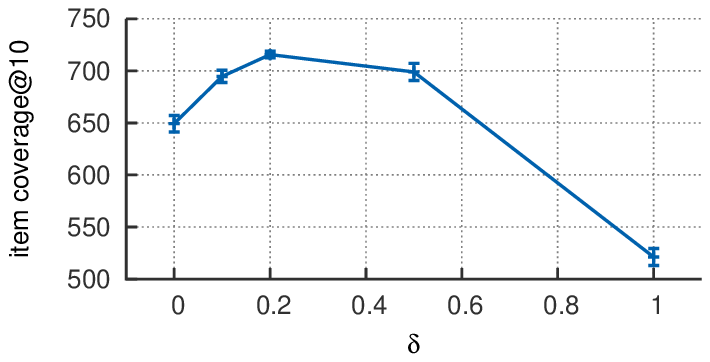}
    \end{subfigure}
  \caption{Influence of the diversity bias on the sps and the item coverage.}
  \label{fig:diversity_bias}
\end{figure}

Figure \ref{fig:diversity_bias} shows that the diversity bias offers an easy way to trade precision for item coverage.
For small values of $\delta$ (under $0.2$) we can significantly increase the item coverage with a small loss in sps.
With larger values of $\delta$, the RNN keeps producing more diverse recommendations, but the precision becomes so low that the item coverage actually decreases (the item coverage is the number of distinct items \emph{correctly} recommended).

\subsection{Extra features}

The input of the RNN used in Section \ref{sec:experiments} consisted only in the sequence of items (in the form of a one hot encoding of the last item at each time step).
However, many datasets possess much richer information and those same information might improve the model discovered by the RNN.
We distinguish three types of extra information: the ones that concern a user (age, sex, etc.), the ones that concern an item (category, price, etc.) and the ones that concern a specific user-item interaction (rating, review, etc.).
Movielens 1M has some of each category: 
\begin{itemize}
    \item users' features: age (approximated by seven age ranges), sex, and occupation (chosen among twenty-one options)
    \item items' features: year of release (we used only the decade) and genre (chosen among eighteen categories)
    \item interactions' features: rating value (ten possible values)
\end{itemize}

We represented each feature with a one hot encoding and added them to the normal input. 
In other words, the input of the RNN that uses the sequence plus, for instance, the users' features, consists in 3736 neurons: 3706 for the one hot encoding of the movie, 7 for the age range, 2 for the sex and 21 for the occupation.

Table \ref{tab:feat_results} shows the influence of using the users' features, the items' features or the interactions' features separately, followed at last by their combined effect.
Interestingly, the gain is only significant when all the extra features are combined, and even then this gain remains small.
This suggests that the sequence of actions already contains most of the information and those extra features are in a way or another already implicit in these sequences.

\begin{table*}[t]
  \caption{Effect of extra features for the RNN on Movielens 1M}
  \label{tab:feat_results}
  \centering
  \begin{tabular}{lllll}
    \toprule
    & \multicolumn{4}{c}{Metrics (@10)} \\
    \cmidrule{2-5}
    Features & sps (\%) & Item coverage & User coverage (\%) & rec (\%) \\
    \midrule
    no extra features & $33.69 \pm 0.58$ & $649.22 \pm 7.88$ & $86.62 \pm 0.38$ & $7.63 \pm 0.1$ \\
    users' features & $33.78 \pm 0.74$ & $656.44 \pm 6.28$ & $87.89 \pm 0.48$ & $7.79 \pm 0.06$ \\
    item's features & $34.07 \pm 0.63$ & $656.67 \pm 5.07$ & $87.9 \pm 0.3$ & $7.69 \pm 0.03$ \\
    interactions' features & $33.04 \pm 0.61$ & $652.56 \pm 9.37$ & $86.89 \pm 0.36$ & $7.56 \pm 0.06$ \\
    all features & $34.97 \pm 0.81$ & $666.17 \pm 3.87$ & $87.33 \pm 0.59$ & $7.87 \pm 0.07$ \\
    \bottomrule
  \end{tabular}
\end{table*}

\section{Related Work}

Deep learning techniques are getting more and more attention in the recommender system community, but we know of only two attempts to use recurrent neural network for collaborative filtering.
Spotify might have been using it as far back as 2014 \cite{spotify_rnn} to build playlists.
They however do not seem to be using gated RNN, and they are using a hierachical softmax output in order to deal with the very large number of items (they have much more songs than netflix or movielens have movies).
More recently, \cite{session_rnn} has applied gated RNN to session based collaborative filtering.
Interestingly, they have used other objective functions than the categorical cross-entropy, namely the Bayesian personalized ranking (the same used by BPR-MF) and an other ranking loss called TOP1 that they devised for the task.

Some earlier works have framed collaborative filtering as a sequence prediction problem and used simpler Markov chain methods to solve it.
In the early 2000s, \cite{MC} used a simple Markov model and tested it for web-page recommendation. \cite{pattern_mining} adopted a similar approach, using sequential pattern mining.
Both showed the superiority of methods based on sequence over nearest-neighbors approaches.
In \cite{MDP1, MDP2}, Brafman et al. defended the view of recommendation systems as a Markov decision process, and although the predictive model was not their main focus, they did present in \cite{MDP1} a Markov chain approach, improved by some heuristics such as skipping and clustering.

More recently, \cite{MCMF} introduced a rather fair approach to build personalized Markov chain, exploiting matrix factorization to fight the sparsity problem.
Their method is mainly designed for the next basket recommendation problem, but it would be of great interest to adapt it for a more general recommendation problem.

\section{Conclusion}

We explored the use of recurrent neural network, and in particular the LSTM, for the collaborative filtering problem.
Using RNNs requires the re-frame collaborative filtering as a sequence prediction problem, and it could lead to richer models, taking the evolution of users' taste into account.
Our experiments showed that the LSTM produces very good results on the Movielens and Netflix datasets, and is especially good in terms of short term prediction and item coverage.

Better performance still could be achieve by designing RNNs specifically for the collaborative filtering task, especially at the level of the objective function, but the fact that standard LSTM works already so well is yet another proof of its ability to tackle general problems.

\subsubsection*{Acknowledgments}

R. Devooght is supported by the Belgian Fonds pour la
Recherche dans l’Industrie et l’Agriculture (FRIA, 1.E041.14).

%
\bibliographystyle{abbrv}
\bibliography{biblio}  
%
%

\end{document}